\documentclass[sigplan,screen]{acmart}

\AtBeginDocument{%
  }

\copyrightyear{2026}
\acmYear{2026}
\setcopyright{cc}
\setcctype{by}
\acmConference[DAC '26]{63rd ACM/IEEE Design Automation Conference}{July 26--29, 2026}{Long Beach, CA, USA}
\acmBooktitle{63rd ACM/IEEE Design Automation Conference (DAC '26), July 26--29, 2026, Long Beach, CA, USA}
\acmDOI{10.1145/3770743.3804357}
\acmISBN{979-8-4007-2254-7/2026/07}

\usepackage{multicol}
\usepackage{multirow}
\usepackage{tabularx}
\usepackage{booktabs}

\newcommand{\ul}[1]{\underline{#1}}

\begin{document}

\title{How Can Reinforcement Learning Achieve Expert-level Placement?}

\author{%
  Ruo-Tong Chen$^{1,2,3}$, Ke Xue$^{1,2}$, Chengrui Gao$^{1,2,3}$, Yunqi Shi$^{1,2}$, Tian Xu$^{1,2}$, Peng Xie$^{1,2}$,\\%
  Siyuan Xu$^{3}$, Mingxuan Yuan$^{3}$, Chao Qian$^{1,2,*}$, Zhi-Hua Zhou$^{1,2}$%
}
\affiliation{%
  \institution{%
    $^{1}$State Key Laboratory of Novel Software Technology, Nanjing University, China\\
    $^{2}$School of Artificial Intelligence, Nanjing University, China\\
    $^{3}$Huawei Noah's Ark Lab, China}
  \country{}}

\renewcommand{\thefootnote}{\fnsymbol{footnote}}
\renewcommand{\shortauthors}{Chen et al.}

\begin{abstract}
Chip placement is a critical step in physical design. While reinforcement learning (RL)-based methods have recently emerged, their training primarily focuses on wirelength optimization, and therefore often fail to achieve expert-quality layouts. We identify the reward design as the primary cause for the performance gap with experts, and instead of formalizing intricate processes, we circumvent this by directly learning from expert layouts to derive a reward model. Our approach starts from the final expert layouts to infer step-by-step expert trajectories. Using these trajectories as demonstrations or preferences, we train a model that captures the latent implicit rewards in expert results. Experiments show that our framework can efficiently learn from even a single design and generalize well to unseen cases.
\end{abstract}

\maketitle

\footnotetext[1]{Corresponding author. E-mail: \texttt{qianc@lamda.nju.edu.cn}.}

\section{Introduction}
Chip Placement plays an important role in physical design, as it determines the physical positions of macros (e.g, memory blocks) and standard cells (e.g., logic gates) on the chip canvas~\cite{wiremask-bbo,kahng2023hier}. To handle the various sizes and numbers between macros and standard cells, placement is usually divided into a macro placement stage, followed by a standard cell placement stage. Macro placement provides a fundamental solution for the subsequent processes (e.g., standard cell placement, clock tree synthesis, and routing), thus having a great impact on the final power, performance, and area (PPA) metrics of the circuit~\cite{agnesina2023autodmp}. 

Given the rapidly growing number of large macros in complex modern chips, macro placement has become increasingly complex and tediously lengthy, making the pursuit of automated algorithms imperative. Reinforcement learning (RL) methods have emerged as particularly promising approaches~\cite{nature-graph,deeppr,lai2022maskplace,xue2024reinforcement,geng2024efficient,goldie2024chip}, which typically formulate the macro placement problem as a Markov decision process (MDP), and learn a policy through interaction with a chip placement environment. 
The adaptability and learning-driven optimization framework of RL offers innovative approaches for macro placement, and has made significant progress, especially in reward modeling: 
AlphaChip~\cite{nature-graph,alphachip} receives no reward feedback until all macros have been placed, making the reward signal sparse for policy training; MaskPlace~\cite{lai2022maskplace} solves this problem by introducing a dense macro half-perimeter wirelength (HPWL) reward. 
However, most of the RL methods primarily focus on wirelength optimization during training, consequently struggling to achieve expert-level placement results and satisfactory PPA metrics.

Automatically achieving expert-level layout remains a longstanding challenge. In current design flows, human experts leverage various forms of domain knowledge to guide macro placement, such as design hierarchy~\cite{kahng2023hier} and dataflow structure~\cite{lin2021dataflow}.
Recently, several automated approaches have been proposed to approximate expert-level placement by incorporating expert prior knowledge, including macro grouping~\cite{macro-grouping-1} and periphery biasing~\cite{incre-macro,xue2024reinforcement}. However, these methods still fall short of fully capturing expert knowledge, as much of the expertise is inherently difficult to formalize and integrate into learning-based frameworks~\cite{macro-grouping-1,kahng2022rtlmp,kahng2023hier}.  For example, macro regularity in terms of dead space management and notch avoidance is crucial but difficult to quantify~\cite{kahng2023hier}. Simplified formulations often fail to accurately describe the complex nature of such regularity.

\begin{figure*}[t!]
\centering
\includegraphics[width=0.98\textwidth]{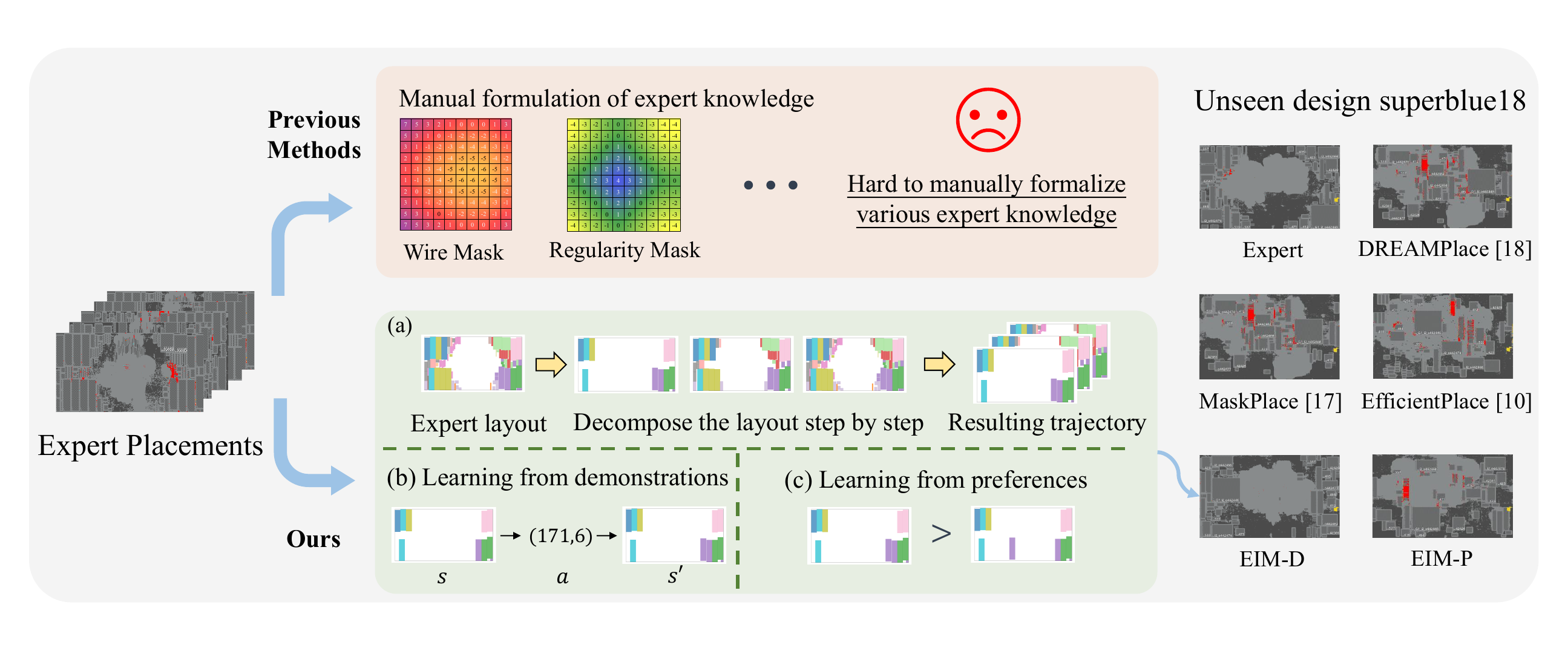} 
\vspace{-5.0mm}
\caption{Illustration of our proposed framework. Instead of manually formalizing intricate expert knowledge, we circumvent this by directly learning from the final expert layouts to derive a reward model. The visualizations of design \texttt{superblue18} of MaskPlace~\cite{lai2022maskplace}, EfficientPlace~\cite{geng2024efficient}, DREAMPlace 4.1.0~\cite{dreamplace4}, EIM-D, EIM-P and Expert are illustrated. The reward models of EIM-D and EIM-P are trained on design \texttt{superblue1}.}
\label{fig:Method-flow}
\vspace{-4.0mm}
\end{figure*}

However, a viable pathway emerges from practical observations: \textit{Even though explicit formalization of expert knowledge remains challenging, numerous expert-level final layout results are readily available.}
Thus, instead of formalizing the detailed expertise, this work circumvents this challenge by learning from the expert-level final layouts themselves. The proposed expert imitation model (EIM) framework starts from the final expert-level layout to infer dense, step-by-step expert trajectories. As shown in Figure~\ref{fig:Method-flow}, EIM first constructs the expert dataset through two key steps: demonstration generation and preference assignment. For demonstration generation, we assume macros are placed sequentially; this allows us to expand a single expert layout into a trajectory of length equal to the number of macros, where each time step yields a transition tuple \((s, a, s')\) that characterizes the incremental placement process. For preference assignment, EIM derives preference signals from these expert transitions: given an expert action $a$ in state $s$, we sample a random legal action \(a_\mathrm{random}\) from the action space to form a preference tuple \((s, a, a_\mathrm{random})\), explicitly encoding the expert’s action preference over alternative choices. Using these trajectories (as demonstrations) and preference tuples (as preference signals), EIM trains a reward model capable of capturing latent implicit rewards through two approaches, i.e., learning from demonstrations (EIM-D) and learning from preferences (EIM-P), respectively.

To comprehensively examine our approach, we conduct experiments on ICCAD 2015 contest benchmark~\cite{iccad15} and OpenROAD-flow-scripts test cases~\cite{ajayi2019toward}. The compared methods in our experiments include two RL methods, where MaskPlace~\cite{lai2022maskplace} as the baseline and  EfficientPlace~\cite{geng2024efficient} as an advanced method, as well as the popular analytical placer DREAMPlace 4.1.0~\cite{lin2020dreamplace,chen2023stronger}. On the seen chip designs, both EIM-D and EIM-P achieve high reward accuracy and the corresponding trained RL policies achieve performance comparable to that of experts. Furthermore, results on unseen chips demonstrate that the learned reward model can still make meaningful progress, validating the generalization capability of our method. Our work preliminarily validates the possibility of learning expert knowledge from final placement outcomes and generalizing to unseen designs, paving the way towards more data-driven and expert-level automated placement with reduced human intervention.

\section{Background}

\textbf{Macro Placement} is a vital stage in modern chip design. The input of a macro placement problem is a netlist $\mathcal{N}=(V,E)$, where $V$ denotes the information (e.g., height and width) about circuit components (including macros and standard cells) for placement, and $E$ is a hyper-graph comprised of hyper-edges $e_i\in E$, which encompasses all components and signifies their interconnectivity during the routing phase. A macro placement method is expected to determine the optimal physical locations of movable macros on the chip canvas such that the objective (e.g., timing metrics) of measuring the chip performance after standard cell placement can be optimized. The mainstream classical approaches for macro placement include the analytical methods~\cite{lu2015eplace,cheng2018replace,lin2020dreamplace,dreamplace4}, which formulate it as a gradient optimization problem, and can place macros and standard cells simultaneously, as well as the black-box optimization methods~\cite{murata1996vlsi,wiremask-bbo}, which optimize the placement layout from a black-box feedback. 

\textbf{Reinforcement Learning for Macro Placement} is a recent popular topic that aims to automatically achieve high-quality placement results~\cite{nature-graph,deeppr,lai2022maskplace,xue2024reinforcement,geng2024efficient,goldie2024chip}. These methods formulate the macro placement process as a MDP, and train an agent to place macros step by step. AlphaChip~\cite{nature-graph} first models macro placement as a RL problem. It divides the chip canvas into several discrete grids, and places one macro in a valid grid coordinate at each time step. However, it suffers from low sample efficiency since no reward is provided until all macros have been placed. To address this issue, MaskPlace~\cite{lai2022maskplace} introduces a dense reward by splitting the calculation of HPWL into cumulative rewards, leading to steadier convergence. To further improve the placement quality, MaskRegulate~\cite{xue2024reinforcement} incorporates a regularity mask to encourage peripheral placements that reflect expert preferences. Existing methods mainly rely on wirelength optimization or manual reward engineering. However, the quality of macro placement is influenced by various factors, such as wirelength, routability, macro halo, power delivery, and electrical characteristics~\cite{lin2019regularity}. How these factors affect the final PPA metrics remains an open problem~\cite{open-problem}.
Therefore, an inadequately crafted reward function may prevent the existing RL-based methods from achieving expert-level results. 
Although it is hard to design an expert-level reward function, it is possible to collect some expert-level placements for imitation.
In this paper, we propose a novel framework that leverages imitation learning to learn an expert-driven reward model, instead of manual reward engineering.

\textbf{Imitation Learning} aims to mimic human expert behavior in a given task, which can be divided into two categories: policy learning and reward learning. 
Policy learning methods such as behavioral cloning~\cite{bc} try to minimize the action differences between the agent policy and the expert, and formulate the imitation problem as regression or classification tasks. 
Reward learning methods~\cite{irl-1,reward-model-survey} first learn a reward model from expert feedback, and then learn the policy by maximizing the learned reward. 
Unlike policy learning methods, which are vulnerable to the accumulated errors from long-horizon planning and distribution shift, reward learning methods can more effectively extract intrinsic knowledge from expert data~\cite{reward-better}, making them better suited for the chip placement problem.

\section{Expert Imitation Model}\label{sec:3}

In this section, we begin by describing the construction of the expert dataset using expert-level placement layouts in Section~\ref{sec:3-1}. We then present two approaches for imitating the expert model: learning from demonstrations (Section~\ref{sec:3-2}) and learning from preferences (Section~\ref{sec:3-3}). Finally, we discuss policy training based on the learned reward model.

\subsection{Dataset Construction}\label{sec:3-1}
Given an expert-level placement result, we first need to process it into expert data that can be used for reward learning. 
This process can be done by demonstration generation and preference assignment.

\subsubsection{Demonstration Generation}
An expert demonstration is a trajectory that involves many transition tuples $(s,a,s')$, where $s$ denotes the current state, $a$ denotes the current action, and $s'$denotes the next state after applying the action. 
In order to decompose the placement layout into a step-by-step demonstration, we assume that the macros are placed one by one in a specific order. Given that assumption, the placement result can be expanded into a trajectory, of which the length is the number of macros placed. At each time step, the expert observes the current layout (i.e., the macros placed currently, denoting the current state $s$), and then decides the location of the next macro (i.e., the expert action $a$). After that, the expert can obtain the updated layout (i.e., the next state $s'$). By doing so, these transition tuples $(s, a, s')$ decompose the placement outcome, therefore enabling reward learning based on demonstrations.

\subsubsection{Preference Assignment}
To enable learning from preference methods, preferences can be induced from expert transition tuples. For example, given an expert transition $(s, a, s')$, the action $a$ adopted by the expert is more preferred than any other action within the action space (i.e., other legal positions for the current macro). Therefore, a random action $a_{\mathrm{random}}$ different from the expert action can be sampled from the action space to form the preference tuple $(s, a, a_{\mathrm{random}})$.

\subsection{Approach 1: Learning from Demonstration}\label{sec:3-2}
Learning from demonstrations~\cite{reward-model-survey} learns the reward model from
expert trajectories. We choose a representative learning method named IQ-Learn~\cite{iq-learn} for implementation. 
IQ-Learn is an inverse reinforcement learning method that starts from the max entropy inverse reinforcement learning with the original objective function:
\begin{equation*}
\begin{split}
    \max_{r \in \mathcal R} \min_{\pi \in \Pi} L(\pi,r) &= \mathbb E_{\rho_E}[r(s,a,s')] - \mathbb E_{\rho_\pi}[r(s,a,s')] \\
    &- H(\pi) - \psi(r),
\end{split}
\end{equation*}
where $\mathcal R$ denotes the reward function set, $\Pi$ denotes the policy set, $\rho_E$ denotes the occupancy measure of the expert, $\rho_\pi$ denotes the occupancy measure of the learning policy, $H(\pi)$ denotes the entropy of learning policy, and $\psi(r)$ is the reward regularizer. This method looks for a reward function that assigns a high reward to the expert and a low reward to others, while the inner loop searches for the best policy for the current reward function. However, the adversarial training between the reward model and policy makes the training unstable. 

To address this issue, IQ-Learn unifies the parameters of the two models using a state-action value function $Q_\theta(s,a)$ parametrized with $\theta$:
\begin{align*}
    \pi(a|s)  &= \frac{1}{Z_s} \exp Q_\theta(s,a), \\
    r(s,a,s') &=  Q_\theta(s,a)-\gamma \cdot \mathbb{E}_{s'\sim \mathcal P(\cdot | s,a)}[V^*(s')],
\end{align*}
where $Z_s=\sum_a\exp Q_\theta(s,a)$ denotes the normalization factor, $\gamma$ denotes the discount factor, $\mathcal P(\cdot | s,a)$ denotes the transition function, and $V^*(s)=\log\sum_a\exp Q_\theta(s,a)$ denotes the state value function in soft Q-learning~\cite{soft-q}.

We provide our expert imitation model by demonstration (EIM-D in short) following this basic framework, and extract the reward model in a transition deterministic manner:
$r(s,a,s') =  Q_\theta(s,a)-\gamma V^*(s'),$
where $s'$ is the next state according to the deterministic transition function $s'=\mathcal P(s,a)$.

\subsection{Approach 2: Learning from Preference}\label{sec:3-3}
Learning from preferences learns the reward model based on expert preferences~\cite{pbrl}. 
We adopt the RLHF~\cite{rlhf} as our reward learning framework, which is a machine learning technique that enables an agent to learn optimal behavior by incorporating human evaluations, such as preferences, ratings, or corrections, into its training process.

The training dataset $D$ of our proposed EIM-P consists of multiple preference tuples $(s,a_c, a_r)$, where $a_c$ denotes the chosen action and $a_r$ denotes the rejected action. This method trains a reward model that assigns a high reward to the chosen action and a low reward to others. To further avoid learned rewards with excessively large scale and complex structure, we add a reward regularization term in the objective function:
\begin{equation*}
    \mathcal J(\theta)=\mathbb E_{(s,a_c, a_r) \sim D}[
    \log \sigma\left(
    r_\theta(s,a_c) - r_\theta(s,a_r)
    \right)
    ] - \psi(r),
\end{equation*}
where $\theta$ denotes the parameter of the reward model $r_\theta$, $\sigma(\cdot)$ denotes the sigmoid function, and $\psi(r)=\alpha \cdot|r|^2$ denotes the reward regularizer with weighted parameter $\alpha$.

\subsection{Policy Training}
After learning reward models from expert layouts, it is essential to select a high-quality reward model for policy training. To facilitate the evaluation of the reward model, we propose a reward accuracy criterion based on the Top-1 accuracy. Specifically,
we first decompose the expert trajectories into $n$ transition tuples $(s, a_e, s')$. 
For each state, we randomly sample $m$ legal actions from the action space to form a validation tuple $(s, \mathcal A_e)$, where $\mathcal A_e = \{a_e, a_1, \cdots, a_m\}$ denotes the set containing the expert action $a_e$ and the $m$ sampled actions. All validation tuples collectively form the validation set $D$. Then, the reward accuracy is computed as follows:
\begin{equation*}
    \text{Acc}(r_\theta)=\frac{1}{n}\sum_{(s,\mathcal A_e) \in D}
    \mathbb I\left(\arg\limits_{a\in \mathcal A_e}\max r_\theta(s,a)  = a_e\right),
\end{equation*}
where $n$ denotes the number of validation tuples, and $\mathbb I(\cdot)$ denotes the indicator function. This metric evaluates whether the expert action receives the highest reward among different actions under each state according to the reward model $r_\theta$. A higher value indicates a better reward quality.

Once the reward model is chosen, we employ a policy network with the same architecture as MaskPlace~\cite{lai2022maskplace}, in which the policy is implemented using a convolutional encoder-decoder network.
The chip canvas is divided into $N \times N$ grids, leading to a discrete action space of size $N^2$.
We train the policy using the proximal policy optimization~\cite{schulman2017proximal} algorithm. 
For a fair comparison, we use the same policy network as MaskPlace~\cite{lai2022maskplace}, where the input has a set of pixel-level feature maps including view mask (describes the overall distribution of chip modules), position mask (describes the available physical positions), and wire mask (describes the incremental value of HPWL), and the output is an action that maps to a valid grid coordinate.

\section{Experiment}
\textbf{Benchmarks.}
To comprehensively examine our method, we use eight designs from the ICCAD 2015 contest C benchmark suite~\cite{iccad15} and six RTL designs from the OpenROAD-flow-scripts~\cite{ajayi2019toward} as our test-beds.

\textbf{Training and Evaluation.}
For the first design in each benchmark (i.e., \texttt{ariane133} from the OpenROAD benchmark and \texttt{superblue1} from the ICCAD 2015 benchmark),  
we obtained 50 placement layouts assisted by the human expert to construct the learning dataset, 
following the procedure described in Section~\ref{sec:3-2}. Subsequently, the reward models are trained using methods EIM-D and EIM-P on \texttt{ariane133} and \texttt{superblue1}, respectively.
In order to validate the quality of learned reward on unseen designs, we further acquire several expert-level placements on designs \texttt{ariane136} and \texttt{superblue18}, and calculate the reward accuracy metric for evaluation. 

\textbf{PPA Evaluation.} To analyze the quality of the learned reward more thoroughly, we further compare the policies trained using the learned reward models. For a fair comparison, the policy architecture follows that of MaskPlace.
For the ICCAD 2015 benchmark, we apply global placement using DREAMPlace~\cite{dreamplace4} for all macro placement methods, and evaluate routing and timing metrics using 
commercial global router.
Several performance indicators are adopted for comprehensive comparison, including routed wirelength (rWL, in m, reported by \texttt{EarlyGlobalRoute}), horizontal/vertical routed overflow (rOH/V, in \%), worst negative slack (WNS, in ns), total negative slack (TNS, in $\times10^5$ns), and number of violated paths (NVP, in $\times 10^4$). For the OpenROAD benchmark, we use the OpenROAD framework to execute the full physical design flow, including global placement, clock tree synthesis, global routing, and detailed routing. To ensure an accurate performance assessment on OpenROAD cases, we report PPA metrics including rWL (m), WNS (ns), TNS (ns), number of global routing overflows (\# overflow), and number of design rule check violations (\# DRC). In each table, the best and runner-up results are \textbf{bolded} and \underline{underlined}, respectively.  

\subsection{Preliminary Studies}

In this section, we first evaluate the learned reward models using the reward accuracy metric. Then, we compare the placement policies with the Expert's layout on the \texttt{ariane133} and \texttt{superblue1} designs, respectively, for further analysis.

The results of reward accuracy are presented in Table~\ref{tab:exp1-reward-learning}.
Both EIM-D and EIM-P demonstrate substantial improvements on \texttt{ariane133} and \texttt{superblue1} compared to the Random Model (column 2). As shown in column 5, EIM-P achieves an accuracy close to 1 on both \texttt{ariane133} and \texttt{superblue1}, indicating its strong ability to imitate expert behavior.
Given the similarity between \texttt{ariane136} and \texttt{ariane133} in terms of macro size and quantity, EIM-P also performs remarkably well on \texttt{ariane136}.
However, its accuracy drops considerably on \texttt{superblue18}, a design that differs significantly from \texttt{superblue1}, revealing the limited generalization capability of EIM-P on previously unseen designs.
Regarding EIM-D (shown in column 4), although its performance on the training set does not match that of EIM-P, it achieves significantly better accuracy on \texttt{superblue18}, demonstrating stronger generalization ability even on significantly different unseen designs.
As shown in column 3, MaskPlace~\cite{lai2022maskplace}, which uses macro HPWL as its reward, aligns well with expert behavior on \texttt{ariane133} and \texttt{ariane136}, but struggles to match the expert on other designs.

The results of policy comparisons are shown in Table~\ref{tab:exp-ppa-iid}. On design \texttt{superblue1}, EIM-D and EIM-P achieve comparable results: EIM-D performs better on routing metrics such as rWL, rOH, and rOV, while EIM-P outperforms in timing metrics, including WNS and TNS. When compared to external methods, both EIM-D and EIM-P outperform MaskPlace, and achieve comparable performance to that of experts. On design \texttt{ariane133}, the results are similar where both EIM-D and EIM-P achieve superior performance. However, although EIM-P achieves the highest reward accuracy, it overfits with the expert data, thus leading to inferior routing and timing metrics to EIM-D.

\subsection{Main Results}

In this section, we further evaluate the generalization capability of the learned reward models by directly applying them to unseen designs. Specifically, the model trained on \texttt{ariane133} is applied to other cases from the OpenROAD benchmark, while the one trained on \texttt{superblue1} is reused on designs from the ICCAD 2015 benchmark, both without any retraining.
To demonstrate the efficiency of our proposed methods, results from EfficientPlace~\cite{geng2024efficient} and DREAMPlace 4.1.0~\cite{dreamplace4} are incorporated for a comprehensive comparison. 

\textbf{Results on ICCAD 2015.} 
As shown in Table~\ref{tab:exp-ppa-ood-iccad},
EIM-D achieves better average ranks than EIM-P across nearly all metrics. Specifically, EIM-D reduces rWL by 5.40\%, and achieves significant reductions in rOH and rOV by 75.55\% and 70.64\%, respectively, averaged across designs. Notably, EIM-D consistently delivers near-optimal congestion results across all test cases, demonstrating its superior performance in terms of routability.
In terms of timing performance, although EIM-D ranks lower in WNS, it achieves better results in TNS and NVP metrics. This outcome aligns with the reward quality comparison shown in Table~\ref{tab:exp1-reward-learning}, where the reward model trained by EIM-D exhibits stronger generalization capability on unseen designs, ultimately enabling better overall policy performance compared to EIM-P.

\begin{table}[t!]
\renewcommand{\arraystretch}{0.9}
\caption{Reward accuracy of our methods on the four designs, where the last two rows are unseen designs during training.}
\vspace{-3mm}
\label{tab:exp1-reward-learning}
\resizebox{0.45\textwidth}{!}{
\begin{tabular}{lcccc}
\toprule \textbf{Benchmark} & \textbf{Random Model} & \textbf{MaskPlace} & \textbf{EIM-D} & \textbf{EIM-P} \\ 
\midrule 
\texttt{ariane133} & 0.02 & 0.54 & \ul{0.77} & \bf{0.99} \\ 
\texttt{superblue1} & 0.04 & 0.12 & \ul{0.78} & \bf{1.00} \\ 
\midrule
\texttt{ariane136}  & 0.07 & 0.45 & \ul{0.64} & \bf{0.99} \\
\texttt{superblue18} & 0.09 & \ul{0.14} & \bf{0.38} & 0.09 \\
\bottomrule
\end{tabular}}
\vspace{-3mm}
\end{table}

\begin{table}[t!]
\centering
\renewcommand{\arraystretch}{0.9}
\caption{PPA results on the two seen designs, where the reward models are trained on the corresponding benchmark.}
\vspace{-3mm}
\label{tab:exp-ppa-iid}
\resizebox{0.95\linewidth}{!}{
\begin{tabular}{l|l|ccccc}
\toprule
\textbf{Benchmark} & \textbf{Metrics} & \textbf{MaskPlace} & \textbf{EIM-D} & \textbf{EIM-P} & \textbf{Expert} \\
\midrule
\multirow{5}{*}{\texttt{ariane133}}
& rWL & \bf{7.21} & 8.91 & 9.16 & \ul{7.71} \\
& WNS & -1.28 & \ul{-1.15} & -1.17 & \bf{-1.04} \\
& TNS & -3760.32 & \ul{-3067.01} & -3146.82 & \bf{-2766.97} \\
& \# overflow & \bf{0} & \bf{0} & \bf{0} & \bf{0} \\
& \# DRC & 20 & \ul{18} & 59 & \bf{17} \\
\midrule
\multirow{6}{*}{\texttt{superblue1}}
& rWL & 139.40 & \ul{121.52} & 132.80 & \bf{108.49} \\
& rOH & 4.61 & \bf{0.17} & 5.01 & \ul{0.28} \\
& rOV & 0.39 & \ul{0.07} & 0.27 & \bf{0.04} \\
& WNS & -77.36 & -77.99 & \bf{-59.18} & \ul{-66.50} \\
& TNS & -3.15 & -1.72 & \ul{-1.53} & \bf{-1.29} \\
& NVP & 2.73 & \ul{1.84} & 2.21 & \bf{1.51} \\
\bottomrule
\end{tabular}
}
\vspace{-4mm}
\end{table}

Among all RL-based methods, EIM-D, EIM-P, and MaskPlace share the same network architecture and training process. Nevertheless, both EIM-D and EIM-P outperform MaskPlace across all evaluated metrics, achieving average improvements of 9.16\% and 3.89\% in rWL, and significantly reducing average congestion by 80.17\% (horizontal) and 24.19\% (vertical), respectively. Similarly, our proposed methods achieve higher average ranks than MaskPlace in WNS and TNS, and reduce NVP by 21.25\% and 14.40\% on average. These results confirm the effectiveness of expert-imitation-based rewards compared to macro HPWL-based rewards. EIM-D generally demonstrates superior overall performance, achieving the highest average ranks in rWL, rOH, rOV, TNS, and NVP. 
Specifically, when compared to the two RL-based methods, EIM-D reduces rWL by 8.23\%, reduces congestion (rOH/V) by 43.35\%, and reduces NVP by 24.68\%. When compared to analytical methods, the improvements are even more significant: rWL reduces by 14.98\%, congestion is reduced by 92.75\%, and NVP decreases by 30.59\%.

\textbf{Results on OpenROAD.} 
Table~\ref{tab:exp-ppa-ood-openroad} summarizes the average rank of each method across \texttt{ariane136}, \texttt{bp}, \texttt{bp\_be}, \texttt{bp\_fe}, and \texttt{swerv\_wrapper} designs.
EIM-D continues to lead across several key metrics, including WNS, TNS, \# overflow, and \#~DRC.
Regarding the rWL metric, the combination of lower density in OpenROAD designs and EIM-D's strategy of utilizing the entire chip canvas may lead to increased wirelength compared to DREAMPlace 4.1.0.

\begin{table*}[htbp]
\centering
\caption{Results of PPA metrics on seven unseen designs of ICCAD 2015 of method MaskPlace~\cite{lai2022maskplace}, EfficientPlace~\cite{geng2024efficient}, DREAMPlace 4.1.0~\cite{dreamplace4}, EIM-D, and EIM-P. The reward models obtained from EIM-D and EIM-P are trained on the \texttt{superblue1}.}
\vspace{-1mm}
\label{tab:exp-ppa-ood-iccad}
\resizebox{1.0\textwidth}{!}{
\begin{tabular}{l|l|ccccccc|c}
\toprule
\textbf{Method} & \textbf{Metric} & \texttt{superblue3} & \texttt{superblue4} & \texttt{superblue5} & \texttt{superblue7} & \texttt{superblue10} & \texttt{superblue16} & \texttt{superblue18} & \textbf{Avg. Rank} \\
\midrule
\multirow{6}{*}{\textbf{MaskPlace}} 
& rWL & 175.17 & 104.83 & 172.49 & 213.98 & 212.14 & 115.40 & 63.79 & 3.71 \\
& rOH & 8.91 & 12.00 & 2.74 & 2.41 & 0.64 & 8.82 & 0.70 & 3.86 \\
& rOV & 0.66 & 0.52 & \ul{0.18} & 0.56 & \ul{0.05} & 0.16 & 0.14 & 3.43 \\
& WNS & -101.68 & -71.93 & -169.07 & -92.96 & \bf{-67.80} & \bf{-50.24} & -35.26 & 3.29 \\
& TNS & -2.27 & -1.69 & -2.21 & -2.44 & \ul{-3.52} & -2.55 & -1.04 & 4.29 \\
& NVP & 1.99 & 1.79 & 2.01 & 3.71 & 1.92 & 2.54 & 2.96 & 3.71 \\
\midrule
\multirow{6}{*}{\textbf{EfficientPlace}} 
& rWL & 179.52 & \bf{99.38} & 181.77 & \bf{190.40} & \ul{205.93} & 113.97 & 69.36 & 3.00 \\
& rOH & 4.70 & \ul{3.96} & 4.07 & \ul{0.39} & \bf{0.23} & \bf{1.09} & 0.56 & \ul{2.29} \\
& rOV & \ul{0.27} & \ul{0.19} & 0.20 & \ul{0.11} & 0.06 & 0.14 & 0.19 & 3.00 \\
& WNS & \ul{-89.93} & -69.21 & \ul{-111.13} & \bf{-54.79} & -71.83 & \ul{-51.61} & -35.08 & \bf{2.29} \\
& TNS & -1.89 & -1.32 & \ul{-1.64} & -1.79 & \bf{-3.42} & -2.12 & -0.48 & \ul{2.71} \\
& NVP & 2.04 & 1.46 & 1.52 & 3.28 & \ul{1.82} & 2.55 & 1.13 & 3.29 \\
\midrule
\multirow{6}{*}{\textbf{DREAMPlace}} 
& rWL & 199.40 & 139.71 & \ul{167.29} & 219.87 & 243.87 & 136.05 & \bf{53.34} & 4.00 \\
& rOH & 15.05 & 18.74 & 5.67 & 6.93 & 3.83 & 16.37 & \ul{0.19} & 4.57 \\
& rOV & 7.40 & 11.60 & 2.18 & 5.24 & 2.10 & 1.93 & \ul{0.02} & 4.57 \\
& WNS & -107.95 & -106.38 & \bf{-71.72} & \ul{-59.58} & -240.30 & -65.48 & \ul{-30.26} & 3.57 \\
& TNS & -2.51 & -2.18 & -1.74 & \bf{-1.37} & -4.10 & \ul{-2.10} & \bf{-0.18} & 2.86 \\
& NVP & \ul{1.98} & 2.69 & 2.05 & 3.81 & 2.45 & 3.75 & \bf{0.50} & 4.00 \\
\midrule
\multirow{6}{*}{\textbf{EIM-D}} 
& rWL & \bf{144.87} & 101.08 & \bf{140.72} & \ul{194.80} & \bf{204.12} & \bf{103.05} & 62.92 & \bf{1.71} \\
& rOH & \bf{0.22} & \bf{1.01} & \bf{0.04} & \bf{0.22} & \ul{0.41} & \ul{3.61} & \bf{0.00} & \bf{1.29} \\
& rOV & \bf{0.04} & \bf{0.06} & \bf{0.02} & \bf{0.03} & \bf{0.03} & \bf{0.08} & \bf{0.01} & \bf{1.00} \\
& WNS & \bf{-78.17} & \ul{-64.11} & -143.06 & -66.20 & -99.54 & -51.62 & -47.57 & 3.00 \\
& TNS & \bf{-1.57} & \ul{-1.28} & \bf{-1.21} & \ul{-1.75} & -4.12 & \bf{-2.07} & -0.31 & \bf{2.00} \\
& NVP & \bf{1.14} & \bf{1.09} & 1.37 & \bf{1.88} & \bf{1.76} & \ul{2.33} & 0.69 & \bf{1.43} \\
\midrule
\multirow{6}{*}{\textbf{EIM-P}} 
& rWL & \ul{157.92} & \ul{99.84} & 168.07 & 213.48 & 218.89 & \ul{107.29} & \ul{59.95} & \ul{2.57} \\
& rOH & \ul{4.24} & 7.58 & \ul{1.86} & 4.02 & 0.50 & 5.73 & 0.57 & 3.00 \\
& rOV & \ul{0.27} & 0.29 & \ul{0.18} & 0.24 & 0.08 & \ul{0.09} & 0.05 & \ul{2.71} \\
& WNS & -100.58 & \bf{-50.36} & -284.17 & -71.43 & \ul{-69.64} & -63.24 & \bf{-21.78} & \ul{2.86} \\
& TNS & \ul{-1.84} & \bf{-1.18} & -2.07 & -2.00 & -4.90 & -2.17 & \ul{-0.23} & 3.14 \\
& NVP & \ul{1.98} & \ul{1.42} & 2.51 & \ul{3.00} & 2.10 & \bf{2.11} & \ul{0.66} & \ul{2.57} \\
\bottomrule
\end{tabular}
}
\vspace{-3mm}
\end{table*}

\begin{table*}[htbp]
\centering
\caption{Results of PPA metrics on five unseen designs of OpenROAD of method MaskPlace~\cite{lai2022maskplace}, EfficientPlace~\cite{geng2024efficient}, DREAMPlace 4.1.0~\cite{dreamplace4}, EIM-D, and EIM-P. The reward models obtained from EIM-D and EIM-P are trained on the \texttt{ariane133}.}
\vspace{-3mm}
\label{tab:exp-ppa-ood-openroad}
\resizebox{0.7\textwidth}{!}{
\begin{tabular}{l|l|cccccc}
\toprule
\textbf{Rank} & \textbf{Metrics} & \textbf{MaskPlace} & \textbf{EfficientPlace} & \textbf{DREAMPlace} & \textbf{EIM-D} & \textbf{EIM-P} \\
\midrule
\multirow{5}{*}{\textbf{Avg. Rank}} 
& rWL & 3.00 & 3.00 & \bf{2.20} & \ul{2.80} & 4.00 \\
& WNS & 3.60 & \ul{2.60} & 2.80 & \bf{2.40} & 3.60 \\
& TNS & 4.20 & 3.40 & \ul{2.40} & \bf{2.00} & 3.00 \\
& \# overflow & \ul{2.80} & 4.00 & 3.00 & \bf{1.20} & 3.40 \\
& \# DRC & \ul{2.40} & 3.80 & 4.00 & \bf{1.40} & 3.40 \\
\bottomrule
\end{tabular}
\vspace{-5.0mm}
}
\end{table*}

\section{Conclusion and Discussion}
In this work, we propose a framework that directly learns an expert-driven reward model from the final expert layouts. By decomposing expert layouts into step-by-step trajectories, we treat these as distinct forms of human feedback, thus introducing two specialized methods: EIM-D and EIM-P. Experimental results on the OpenROAD and the ICCAD benchmarks show that our proposed framework can efficiently learn from even a single design and generalize well to unseen cases, outperforming the state-of-the-art RL-based methods and analytical-based methods.
EIM exemplifies a hybrid approach that integrates domain knowledge with data-driven learning, which has been a promising direction in AI research such as abductive learning~\citep{abductive}.
Future research will explore dataset augmentation through synthetic data to alleviate the scarcity of expert data, and integrate advanced techniques like adversarial inverse RL and diffusion model-based imitation learning for more sophisticated reward modeling.

\section*{Acknowledgment}
This work was supported by the National Science Foundation of China (624B2069), 
the Jiangsu Science Foundation Leading-edge Technology Program (BK20232003),
the Fundamental Research Funds for the Central Universities (14380020), and the Fundamental and Interdisciplinary Disciplines Breakthrough Plan of the Ministry of Education of China (No. JYB2025XDXM118).

\clearpage
\clearpage
\newpage
\bibliographystyle{ACM-Reference-Format}
\bibliography{main}

\end{document}